\documentclass[twocolumn,showpacs,preprintnumbers,amsmath,amssymb]{revtex4}

\usepackage{graphicx}
\usepackage{dcolumn}
\usepackage{bm}

\begin{document}


\title{Fluorescence measurements of expanding strongly-coupled neutral plasmas}

\author{E. A. Cummings}
 \altaffiliation[Present Address: ]{Lockheed Martin Space Systems Company, Sunnyvale,
 CA 94089}
\author{J. E. Daily, D. S. Durfee, and S. D. Bergeson}%
 \email{scott.bergeson@byu.edu}
 \affiliation{Brigham Young University, Department of Physics and
 Astronomy, Provo, UT 84602}

\date{\today}

\begin{abstract}
We report new detailed density profile measurements in expanding
strongly-coupled neutral plasmas.  Using laser-induced
fluorescence techniques, we determine plasma densities in the
range of $10^{5}$ to $10^{9}$ cm$^{-3}$ with a time resolution
limit as small as 7 ns.  Strong-coupling in the plasma ions is
inferred directly from the fluorescence signals.  Evidence for
strong-coupling at late times is presented, confirming a recent
theoretical result.
\end{abstract}

\maketitle

Strongly-coupled Coulomb systems appear in extreme conditions,
such as in quark-gluon plasmas \cite{shuryak04}, in laser-fusion
plasmas \cite{woolsey98}, and in some astrophysical settings.  A
new class of strongly-interacting neutral plasmas was recently
demonstrated using the tools of laser-cooling and trapping
\cite{killian99,kulin00,killian01,simien04,chen04}. These
``ultracold'' neutral plasmas occupy a unique position in phase
space. In these plasmas it is possible to create
strongly-interacting Coulomb systems at modest densities because
the initial electron and ion temperatures can be in the
milliKelvin range.  The initial ion-ion and ion-electron
interaction strength can also be selected with great precision.

Recent experimental work in this field has used absorption imaging
techniques to make temperature and density measurements in
expanding ultracold neutral plasmas \cite{simien04,chen04}.  This
work explored the 50 to 1000 ns time period after plasma formation
in great detail. Correlation-induced heating was observed in the
plasma ions. The ion coupling parameter, given as the ratio of
nearest-neighbor Coulomb energy to the kinetic energy,
equilibrated just inside the strongly-coupled regime, with a
coupling parameter around 2.

Radio-frequency (RF) excitation techniques have also been used to
determine the average ion density and the electron temperature in
these systems \cite{kulin00,bergeson03,roberts04}.  These studies
confirm theoretical predictions regarding the generally
self-similar Gaussian expansion of the ions and the clamping of
the electron temperature in the weakly-coupled regime.

In this letter we report laser-induced-fluorescence measurements
of ions in expanding strongly-coupled plasmas as a tool to study
the spatial and temporal evolution of the ion temperature and
density. This measurement technique has a 7 ns temporal resolution
limit.  We measure plasma densities as low as $10^{5}$ cm$^{-3}$
at effective plasma temperatures of 100 K. The maximum density
that can be measured is limited by radiation trapping, and for
spherically-symmetric systems in the milliKelvin range is limited
to around $10^9$ cm$^{-3}$.  The temporal resolution and dynamic
range of this method in ultracold plasma measurements surpass
those currently seen in absorption spectroscopy, and rival the
sensitivity of RF spectroscopic methods.

Much of the experimental setup has been described previously
\cite{ludlow01}. We create a calcium magneto-optical trap (MOT) using
the resonance transition at 423 nm.  Up to 50 mW of 423 nm
radiation is generated by frequency-doubling a diode laser system
at 846 nm using a periodically-poled KTP crysal in a resonant
build-up cavity. The 423 nm MOT light is detuned one line-width
below the atomic transition, making the MOT temperature around 1
mK.  The density distribution is approximately Gaussian in $x, y,$
and $z$, of the form
\begin{equation}
n=n_0 \mbox{exp}[-(x^2+y^2)/\sigma_0^2
-z^2/\beta_0^2]. \label{eqn:dist}
\end{equation}
\noindent The $1/e^2$ radius is $\sigma_0=0.5$ mm in the $x$ and $y$
dimensions, and $\beta_0 = 2.5 \sigma_0 = 1.25$ mm in
the $z$ dimension.   We use a grating-stabilized 657 nm diode laser as a
re-pumper to increase the number of atoms in the trap. The peak density
determined by absorption measurements on the atoms is $4 \times 10^{9}$
cm$^{-3}$.

We photo-ionize the atoms in the MOT using a two-color, two-photon
ionization process.  A portion of the 846 nm diode laser radiation
from the MOT laser is pulse-amplified in a pair of YAG-pumped dye
cells and frequency doubled.  This produces a 3 ns-duration laser
pulse at 423 nm with a pulse energy around 1 $\mu$J.  This laser
pulse passes through the MOT, and its peak intensity is a few thousand
times greater than the saturation intensity.  A second YAG-pumped dye laser at
390 nm counter-propagates the 423 nm pulse and  excites the MOT
atoms to low-energy states above the ionization potential.  We
photo-ionize 85-90\% of the ground-state atoms in the MOT.  The
minimum electron temperature is limited by the 30 GHz (measured)
bandwidth of the 390 nm laser to about 1 K.

\begin{figure}
\includegraphics[width=3.3in]{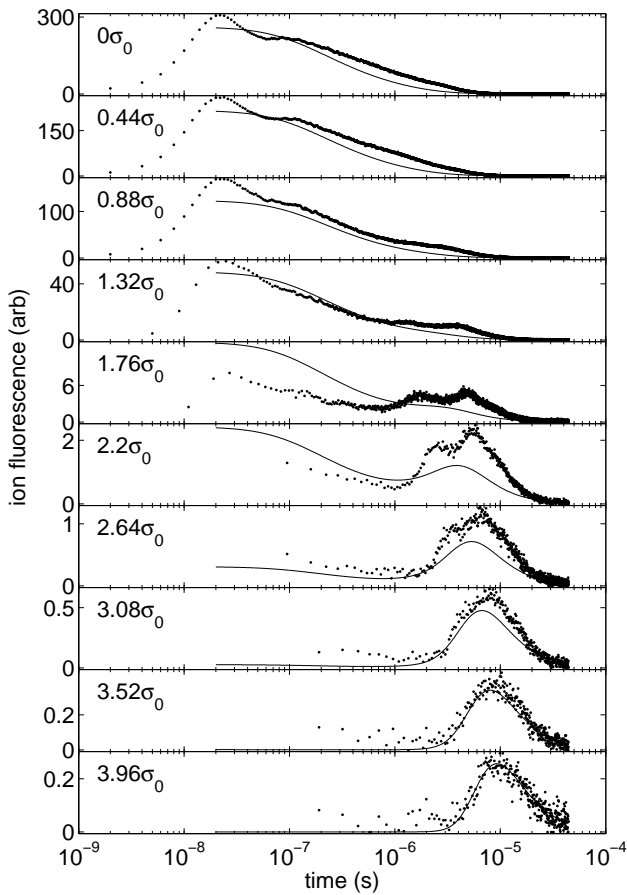}
\caption{Laser-induced-fluorescence of ions in an expanding
ultracold plasma.  In the upper panel, the probe beam is parallel
to (and collinear with) the long axis of the initial plasma
density distribution. In each successive panel, the probe beam is
moved 0.22 mm = 0.44$\sigma_0$ farther away from this symmetry
axis, where $\sigma_0$ is the 1/e$^2$ transverse Gaussian
radius of the initial
plasma density
distribution in the $x$ dimension.
Note that the time scale is logarithmic and the fluorescence
scale is linear.  The solid line is the expected signal from a thermal
expansion model, as explained in the text.
\label{fig:ionExpansion2}}
\end{figure}

Ions in the plasma scatter light from a probe laser beam tuned to
the Ca {\sc II} $^2S_{1/2} - ^2P_{1/2}$ transition at 397 nm.  
The probe laser is generated by a grating-stabilized violet diode laser 
locked to the calcium ion transition using the DAVLL technique \cite{corwin98} 
in a large-bore, low-pressure hollow cathode discharge of our
own design. 
The probe laser is spatially filtered and a few $\mu$W of power is
focused to a Gaussian waist of 130 $\mu$m in the
MOT.  We average repeated measurements of the scattered 397 nm
radiation with the probe laser in a given position.  This produces
a time-resolved signal proportional to the number of atoms
resonant with the probe beam in a particular column of the plasma.
By translating a mirror just outside the MOT chamber, we scan
the probe laser across the ion cloud.  In this manner we obtain
temporal and spatial information about the plasma expansion.
Typical fluorescence measurements are shown in Fig.
\ref{fig:ionExpansion2}.

When the probe beam is on axis (Fig. \ref{fig:ionExpansion2}, top
panel), the fluorescence signal takes roughly 20 ns to reach its
maximum value.  This duration is a few times the 7 ns radiative
lifetime of the Ca {\sc II} $^2P_{1/2}$ level.  This is an
example of the classic Rabi two-level atom problem with
spontaneous emission, where a collection of atoms driven on
resonance well below the saturation intensity will reach a steady
excited-state population after a few radiative lifetimes.  We also see
this same turn-on time when we perform absorption measurements on
the plasma ions.  A higher probe laser intensity would drive the ion
population into the excited state more quickly, making it possible
to probe the plasma at earlier times.

After the signal peaks, it rapidly decays.  At $t\sim 100$ ns the
decay slows down.  This is due to correlation-induced heating in
the ions, similar to previously published absorption measurements
\cite{chen04}. During the first microsecond of the decay, while
the spatial distribution of the plasma ions has not changed, it is
straightforward to convert this fluorescence decay signal into
velocity as a function of time.  With a few assumptions, we
can determine the temperature, density, and strong-coupling
parameter of the plasma ions.

An atom moving with velocity $v_z$ parallel to the propagation
direction of the probe laser beam has a Lorentzian probability of
scattering a photon, proportional to $[({
v_z }/{ b })^2 +1]^{-1}$.  The constant $b= \gamma\lambda/2$ is
the velocity that corresponds to a Doppler shift equal to the
natural
linewidth, $\gamma=1/2\pi\tau$ is the natural linewidth of the
transition, and $\lambda$ is the transition wavelength. The
initial velocity distribution is Maxwellian, of the form $\exp
(-v_z^2/2v_{th}^2)/\sqrt{\pi}v_{th}$. For systems in thermal
equilibrium, $v_{th}$ is the thermal velocity. As mentioned in the
literature \cite{chen04}, the temperature should be a measure
of the random motion of ions.  In these kinds of
measurements for expanding ultracold plasmas, this random motion
is observable only at early times.  The random motion is
quickly overwhelmed by the directed motion of the plasma expansion.
So $v_{th}$ quickly looses its meaning in a thermodynamic sense.
However, when the ions are ``heated'' by correlations the
Maxwellian approximation is still valid.

\begin{figure}
\includegraphics[width=3.3in]{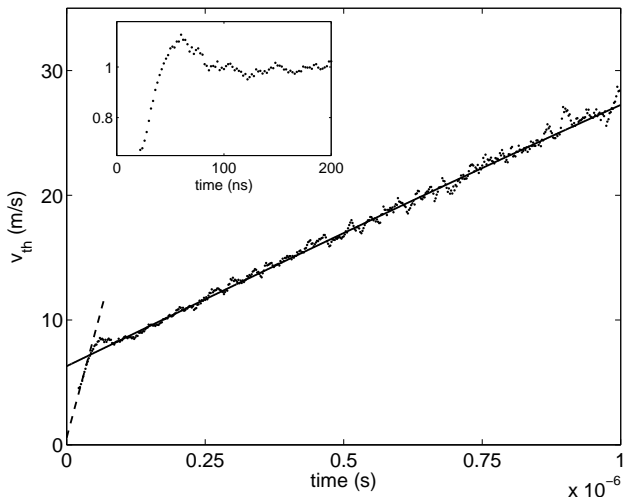}
\caption{The mean $z$-component of the ion velocity as a function
of time after plasma formation.  The points show the velocity
extracted from the ion fluorescence signal shown in the top panel of
Fig. \ref{fig:ionExpansion2}.  The solid line is a
fit of the data to a thermal model \cite{robicheaux03}, as
described in the text. The dashed line shows the extrapolation to
56 cm/s at time $t=0$.  The inset shows the ratio of the data to
the thermal model, accentuating the correlation-induced heating
signal.\label{fig:ionVelocity}}
\end{figure}

We integrate the product of the Lorentzian line shape and the
Maxwellian distribution to get an expression for how the
fluorescence signal, $s(t)$, changes with the width of the ion
velocity distribution as a function of time:
\begin{equation}
s(t) \propto \frac{1}{v_{th}} \exp\left(
\frac{b^2}{2v_{th}^2}\right)
\mbox{erfc}\left(\frac{b}{\sqrt{2}v_{th}}\right) \label{eqn:vel}
\end{equation}
\noindent where erfc$(x)$ is the complimentary error function of
the parameter $x$.  If $v_{th}$ is known at a particular time,
this equation can be inverted to give the change in the thermal
velocity as a function of time.  The initial neutral atom velocity
is near the Doppler limit at 40
cm/s.  The neutral atoms also experience an ionization recoil
velocity of 40 cm/s.  In the inversion of Eq. \ref{eqn:vel}, we
set the scale so that a linear extrapolation of the signal to time
$t=0$ gives the quadrature sum of the Doppler and recoil
velocities, 56 cm/s, as shown in Fig. \ref{fig:ionVelocity}.

A thermal model of ultracold plasma expansions was presented in
Ref. \cite{robicheaux02,robicheaux03}.  We have analytically
solved the coupled differential equations presented in that work
\cite{bergeson03}, and find the analytical solution to the
directed expansion velocity to be
\begin{equation}
v(r,t) = \frac{v_{e}^2t\;r}{\sigma^2(t)} \label{eqn:velocity}
\end{equation}
\noindent where $\sigma(t) = \sqrt{\sigma_0^2 + v_{e}^2 t^2}$ is
the time-dependent width of a self-similar Gaussian expansion.  At
early times, when $v_e^2 t^2 \ll \sigma_0^2$, the velocity
reduces to $v(r,t) \approx v_e^2 t \; r/ \sigma_0^2$.

In applying this expression for the expansion velocity to the data
in Fig. \ref{fig:ionVelocity}, some care is required.  The model
assumes ions initially at rest.  However, in our experiment
the plasma ions have an initial
electrical potential energy which is quickly converted to kinetic
energy (see the discussion of correlation heating below).  This
gives an offset to the $t=0$ velocity predicted by the model.

Otherwise, the application is straightforward.  The probe laser
interacts with ions in a particular volume.  During the first
microsecond, the large-scale ion distribution does not change, even
though the ions are accelerating.
The expression in Eq. \ref{eqn:velocity} predicts
a velocity that depends linearly in time, with $v_e^2 r /\sigma_0^2$ averaged
over the part of the density distribution interacting with the
probe laser beam.  A
fit of our data to this model is shown in Fig
\ref{fig:ionVelocity}.  For the early expansion times, the
lack of spherical symmetry in our
plasma is unimportant in fitting this data, because the average
$v_e^2 r/\sigma_0^2$ appears as a fit parameter.

This figure illustrates correlation-induced heating, and provides
an independent check of the plasma density.  When the plasma is
created, the ions quickly move from their initial disordered state
into a state with greater order.  This motion is initially
synchronized at the plasma frequency.  But because of variations
in the local electric field at each ion, the oscillatory motion
becomes un-synchronized.  The time from plasma
creation to the maximum temperature (minimum signal) in this
oscillation is equal to $t=1/4\omega_p$, where $\omega_p$ is the
average plasma frequency.  As seen in the inset to Fig.
\ref{fig:ionVelocity}, the peak occurs at 60 ns, giving an average
plasma density of $4.0\times 10^8$ cm$^{-3}$, about a factor of
three lower than what we expect based on the ionization fraction
of the neutral MOT.  The cause of this discrepancy is not
readily apparent.  Our numerical modeling of the plasma oscillation
dynamics indicates that the average density determined from the
observed oscillation period reproduces the average plasma
density to within 20\%.  Perhaps this discrepancy indicates
the number of plasma ions that have recombined
into high-lying Rydberg states at early times.

The strong-coupling constant in this system can be determined from
data extracted from Fig. \ref{fig:ionVelocity}. The kinetic energy
of the system is equal to $mv^2/2$, where
$v^2=v_x^2+v_y^2+v_z^2$, and the random thermal motion due to
correlation heating is assumed to be isotropic.  The temperature is
determined from $3k_bT/2 = mv^2/2$.  Using the derived
values of $v_z = 8$ m/s and $n=4\times 10^8$ cm$^{-3}$, the
strong-coupling parameter is $\Gamma = (e^2/4\pi\epsilon_0 k_B
T)(4\pi n/3)^{1/3} = 4$.

During the time from 1 to 40 $\mu$s, the fluorescence signal
depends on the time-evolution of the ion temperature, the overall
plasma expansion velocity, and the density.  As the plasma
expands, the number of ions in the probe laser beam decreases.
Also, as the velocity distribution widens, the fraction of atoms
in the beam that are in resonance with the probe laser frequency
also decreases.

The thermal expansion model in Ref. \cite{robicheaux03} provides a
theoretical framework for interpreting our data.
Most of the ions in the plasma lie in a region in which the electric
field depends linearly on the spatial coordinate.  The ansatz
presented in Ref. \cite{robicheaux02} simplifies the
system by imposing this linear dependence of the electric field
for the entire system.  While this is clearly not true for ions
near the edge of the plasma, the number of ions in that region is
small.  Sophisticated
approaches to modeling this system have shown that the thermal
model captures the major features of the expansion dynamics
\cite{robicheaux02,robicheaux03,pohl04}.

The lack of spherical symmetry in our experiments
changes the slope of the electric field in the $z$ direction
relative to $x$ (or $y$), because the plasma distribution is
elongated in $z$ by a factor of 2.5 (see Eq. \ref{eqn:dist}).
Keeping with the ansatz of Ref. \cite{robicheaux02}, we
assume that the Gaussian plasma expands in a self-similar
fashion.  Similar to Eq. \ref{eqn:velocity}, the expansion
velocity in the $x$ direction is
$v_x(t) = v_{e}^2t\;x/\sigma^2(t)$.  The
expansion velocity in the $z$ direction is
$v_z(t) = v_{e}^2t\;z/\beta^2(t)$, where
$\beta=\sqrt{(2.5\sigma_0)^2 + v_{e}^2 t^2}$.  This result, which
relates the velocity to the position, will be important in
modeling the fluorescence signals, as discussed below.

In the experiment, the probe laser beam is spatially filtered and
focused into the plasma with a Gaussian beam waist of $w=130$
$\mu$m. The position offset of this probe laser relative to the
plasma is denoted by the distance $a$.  The two-dimensional
convolution of the probe laser and the plasma distribution is
proportional to the number of plasma ions per unit length
along the probe beam,
\begin{equation}
\eta=\frac{w^2}{\beta(w^2+2\sigma^2)} \exp\left(
-\frac{2a^2}{w^2+2\sigma^2} \right) \exp\left(
-\frac{z^2}{\beta^2} \right) .
\label{eqn:eta}
\end{equation}
\noindent Multiplying this
expression by the Lorentzian lineshape gives
a probability that an ion in the probe beam column moving with a
$z$-component of velocity $v_z$ interacts with the probe laser.
We substitute for $v_z$ the expression derived previously,
and integrate Eq. \ref{eqn:eta} over $z$ to derive
the predicted time-dependence of the
fluorescence signal:
\begin{equation}
s(t) \propto \mbox{erfc}\left(\frac{\ell}{\beta}\right)
 \frac{w^2 \ell}{\beta(w^2+2\sigma^2)}
 \exp\left(-\frac{2a^2}{w^2+2\sigma^2} +
 \frac{\ell^2}{\beta^2}\right)
\end{equation}
\noindent where $\ell \equiv \gamma \lambda \sigma^2(t) / 2 t
v_{e}^2$.  This derivation neglects the influence
of the thermal velocity, and is valid at times where the directed
expansion velocity dominates the thermal velocity (a regime
opposite to that for Eq. \ref{eqn:vel}).  Adding the thermal
velocity to the model would further reduce the predicted signal
levels at early times.

This fluorescence model is the solid line plotted with the data in
Fig. \ref{fig:ionExpansion2}. The only two adjustable parameters
are the expansion velocity $v_{e}$ and the overall
amplitude.  They are chosen so that the model
agrees with the data in the last panel of the figure. Considering
the simplicity of this model, it is surprising that it matches the
general shape and order of magnitude of the fluorescence signal.

Relative to the model, the experimental data are consistently
higher at later times.  Additionally, data recorded with
the probe laser beam between $a=\sigma_0$ and $3\sigma_0$
show structure that cannot be explained by the model.
Predictions of ion-acoustic waves freezing into the plasma
expansion have been made.  However, the observed modulations are
not stationary relative to the plasma expansion.  A vertical slice
through all of the panels in Fig. \ref{fig:ionExpansion2} gives
the density profile at a given time. The data show that two local
maxima that develop in the interior of the plasma coalesce into
one another and disappear relative to the smooth Gaussian shape
predicted by the model (lowest frame). Contrary to predictions,
we see no evidence for shock formation in these
plasmas.

It is likely that the plasma ions are strongly coupled throughout
the expansion.  The ions equilibrate
after 100 ns with a coupling parameter of $\Gamma = 4$. Recent
hybrid molecular dynamics calculations
\cite{pohl05} show that after an initial oscillation, the coupling
parameter always increases. The absence of a shock front in the
plasma expansion and also the observed suppression of local
density variations is evidence that the plasma ion system is
collisionally stiff, once again suggesting strong coupling.

In conclusion, we have demonstrated a new technique for
determining the the density of ions in an expanding ultracold
plasma using fluorescence spectroscopy.  Our implementation of
this technique has a time resolution of 7 ns.  We report
measurements of the density profile for up to 50 $\mu$s, and
demonstrate a sensitivity limit around 10$^5$ cm$^{-3}$.  This
method surpasses the time resolution and dynamic range  of
previously reported techniques.  It rivals the sensitivity limit
of RF techniques, and provides detailed spatial information of the
plasma density profile.  We have created a plasma just inside the
strongly-coupled regime, and observed changes in the plasma density
over time. Based on recent theoretical work, it is likely that
these plasmas are strongly-coupled for the entire measured
expansion. The general features of our measurements are in
reasonable agreement with a thermal model, although the remaining
discrepancies require more sophisticated methods to interpret
quantitatively.

This research is supported in part by Brigham Young University, the
Research Corporation, and the National Science Foundation (Grant No.
PHY-9985027).


\begin{thebibliography}{99}

\bibitem{shuryak04}
Edward V. Shuryak and Ismail Zahed, Phys. Rev. C 70, 021901 (2004)

\bibitem{woolsey98}
Nigel C. Woolsey, David Riley, and Eran Nardi, Rev. Sci. Instrum.
69, 418 (1998)

\bibitem{killian99}
T. C. Killian, S. Kulin, S. D. Bergeson, L. A. Orozco, C. Orzel,
and S. L. Rolston Phys. Rev. Lett. 83, 4776 (1999)

\bibitem{kulin00}
S. Kulin, T. C. Killian, S. D. Bergeson, and S. L. Rolston
Phys. Rev. Lett. 85, 318 (2000)

\bibitem{killian01}
T. C. Killian, M. J. Lim, S. Kulin, R. Dumke, S. D. Bergeson, and
S. L. Rolston Phys. Rev. Lett. 86, 3759 (2001)

\bibitem{simien04}
C. E. Simien, Y. C. Chen, P. Gupta, S. Laha, Y. N. Martinez, P.
G. Mickelson, S. B. Nagel, and T. C. Killian Phys. Rev. Lett. 92,
143001 (2004)

\bibitem{chen04}
Y. C. Chen, C. E. Simien, S. Laha, P. Gupta, Y. N. Martinez, P.
G. Mickelson, S. B. Nagel, and T. C. Killian Phys. Rev. Lett. 93,
265003 (2004)

\bibitem{bergeson03}
S. D. Bergeson and R. L. Spencer Phys. Rev. E 67, 026414 (2003)

\bibitem{roberts04}
J. L. Roberts, C. D. Fertig, M. J. Lim, and S. L. Rolston, Phys.
Rev. Lett. 92, 253003 (2004).

\bibitem{ludlow01}
A. D. Ludlow, H. M. Nelson, and S. D. Bergeson J. Opt. Soc. Am. B
18, 1813 (2001)

\bibitem{corwin98}
Kristan L. Corwin, Zheng-Tian Lu, Carter F. Hand, Ryan J. Epstein,
Carl E. Wieman, Appl. Opt. 37, 3295 (1998)

\bibitem{robicheaux02}
F. Robicheaux and James D. Hanson Phys. Rev. Lett. 88, 055002
(2002)

\bibitem{robicheaux03}
F. Robicheaux and James D. Hanson Phys. Plasmas 10, 2217 (2003)

\bibitem{pohl04}
T. Pohl, T. Pattard, and J. M. Rost, Phys. Rev. A 70, 033416 (2004)

\bibitem{pohl05}
T. Pohl, T. Pattard, and J. M. Rost,
http://arxiv.org/abs/physics/0503008 (2005)

\bibitem{kuzmin02a}
S. G. Kuzmin and T. M. O'Neil,
Phys. Rev. Lett. 88, 065003 (2002)

\bibitem{kuzmin02b}
S. G. Kuzmin and T. M. O'Neil,
Phys. Plasmas 9, 3743 (2002)

\end{thebibliography}
\end{document}